\documentclass[12pt]{article}
\usepackage{amsfonts,amssymb,amsmath}
\usepackage{graphicx}
\def\s{\sigma}
\def\g{\gamma}

\newtheorem{theorem}{Theorem} 
\newtheorem{dfnt}{Definition}

\newtheorem{prop}{Proposition}
\newtheorem{lemma}{Lemma}
\newtheorem{eg}{Example}
\newtheorem{rmk}{Remark}

\newcommand{\rf}[1]{{\rm \ref{#1}}}

\title{Correlation Inequalities for Generalized Potts Model: General Griffiths' Inequalities}

\author{Nasir Ganikhodjaev$ ^{1,2}$ and Fatimah Abdul Razak $^1$}
\date{}

\begin{document}

\maketitle

\begin{center}
$^1$ Department of Computational and Theoretical Sciences,
Faculty of Science, IIUM, 25200 Kuantan, Malaysia.\\
$^2$ Department of Mechanics and Mathematics, National University of Uzbekistan, Vuzgorodok, 700095, Tashkent, Uzbekistan.\\ 
\vskip 0.1cm

gnasir@iiu.edu.my, fatimah84@gmail.com.
\end{center}

\begin{abstract}
In this paper, correlation inequalities which have been considered on Ising model are extended to q-Potts model. It is considered on generalized Potts model with interaction of any number of spins. We replace the set of spin values $F=\{1,2,\cdots ,q\}$ by the centered set $F=\{-(q-1)/2,-(q-3)/2,\cdots ,(q-3)/2,(q-1)/2\}$. Let $N$ be the subset of one-dimensional lattice with $n$ vertices, $\g=(\s_1,\s_2,\cdots,\s_n):N \rightarrow F^c$ be a configuration where ${(\s_i)}_\g$ is the number which appears as the ith spin (component) in $\g$ and $\s_i$ be a random variable whose value at $\g$ is ${(\s_i)}_\g$. Define $\s^R=\prod_{i \in R}\s_i$ for any list $R$ where any $i \in R$ implies that $i \in N$. We first prove that $\langle \s^R \rangle \ge 0$ then we prove that for any two lists $R$ and $S$, we have $\langle \s^R \s^S \rangle- \langle \s^R \rangle \langle \s^S \rangle \ge 0$. \\


\end{abstract}

\section{Introduction}

Statistical physics seeks to explain the macroscopic behaviour of matter on the basis of its microscopic structure. This includes the analysis of simplified mathematical models \cite{Ge}. Ferromagnetic metal can be regarded as being composed of elementary magnetic moments called spins which are arranged on the vertices of a crystal lattice. The orientation of each spin is random but subject to spin-spin interaction which favors their alignment.

The Potts model \cite{Po} was introduced as a generalization of the Ising model to more than
two components (spins). Ising model considered only up and down spins \cite{KS} whereas Potts model incorporates more possibilities of spins and their interactions. The Potts model describes an easily defined class of statistical mechanics models. At the same time, its rich structure is surprisingly capable of illustrating almost every conceivable nuance of the subject \cite{jv}. The Potts model encompasses a number of problems in statistical physics (see, e.g. \cite{W}).

Griffiths' inequalities  \cite{gr} exhibit the monotonic behaviour of the moments (correlations) in a ferromagnetic Ising system as a function of interactions \cite{KS}. By proving that these correlation inequalities can be applied to Potts model, calculations regarding interactions of q-spins will be simplified and more of its properties can be explained mathematically.

\section{Preliminaries}

Let $N$ denote the index set $\{1, 2, \cdots ,n \}$, consider the space of all spin configurations $(\s_1,\s_2,..,\s_n)$ where each $\s_i$ is allowed the values from $1$ to $q \ (q \ge 2)$. A general configuration is denoted by $\g$ and $(\s_i)_\g$ is the number of values $(1, \cdots ,q)$ which appears as the $i$th spin (component) in $\g$. Let $\Omega$ be the set of all possible configurations.

For each pair $(i,j)$ of distinct indices in $N$ the extended real number
\begin{equation}\label{1}
J_{ij} = J_{ji}\ge 0
\end{equation}
is given $(J_{ji} = \infty$ is permitted). The requirement $J_{ji}\ge 0$ is that the system be ferromagnetic. 
The Hamiltonian of the Potts model is the real valued function on configurations, whose value at the configuration $\g$ is
\begin{equation}
H_\g=-\sum_{1 \leq i < j \leq n} J_{ij}\delta_{(\s_i)_\g(\s_{j})_\g}
\end{equation} 
where $\delta$ is the Kronecer's symbol defined as
\begin{equation*}
\delta_{(\s_i)_\g(\s_{j})_\g}=\left\{
\begin{array}{ll}
1&  if \  (\s_i)_\g=(\s_j)_\g,\\
0&  otherwise.
\end{array}\right.
\end{equation*}

The Gibbs probability $P$ on the space of configurations $\Omega$ is defined by
\begin{equation}
P(\g)=Z^{-1}\exp{(-\beta H_\g)}, \end{equation}
where
\begin{equation}
Z=\sum_\g \exp{(-\beta H_\g)}  \end{equation}
and
\begin{equation}\label{5}
\beta=(kT)^{-1} > 0,  \end{equation}
where $k$ is the Boltzman's constant and $T$ is the absolute temperature. For brevity, $\beta$ will be assumed to be 1 for the rest of the paper which gives the probability as
\begin{equation}
P(\g)=Z^{-1}\exp{(H_\g)}. \end{equation}

The expected value of a random variable $X$ on this probability space $(\Omega,P)$ is called its thermal average and is denoted by angular brackets:
\begin{equation}
\langle X \rangle=E(X)=\sum_\g X(\g)P(\g). \end{equation}


\section{Centered Random Variables}

Let $\s_i$ denote the random variable whose value at $\g$ is $(\s_i)_\g$, that is, it's range is the following set $F=\{1,2,\cdots,q\}$,  then
\begin{equation}
\langle \s_i \rangle=\sum_\g (\s_i)_\g P(\g). \end{equation}
We introduce centered random variable $\s_i^\prime$ whose values are derived from  $\s_i$ such as
\begin{equation} \label{cv}
\s_i^\prime=\s_i-\langle \s_i \rangle .
\end{equation}

\begin{prop} \label{prop1}
For any given $q$ and arbitrary $i \in N$, range $F^c$ of the centered random varibles $\s_i^\prime$ is the following set: \\
\indent i) if $q$ is an odd positive integer, that is, $q=2m+1$, then
\begin{equation*}
F^c=\{-m,-(m-1),\cdots,-1,0,1,\cdots,m-1,m\}; 
\end{equation*}
\indent ii) if q is an even number,that is, $q=2m$, then
\begin{equation*}
F^c=\{-[\frac{\rm(2m-1)}{\rm2}],-[\frac{\rm(2m-3)}{\rm2}],\cdots,-\frac{\rm1}{\rm2},\frac{\rm1}{\rm2},\cdots,[\frac{\rm(2m-3)}{\rm2}],[\frac{\rm(2m-1)}{\rm2}]\}. 
\end{equation*}
\end{prop}
{\it Proof.} To find $\langle \s_i \rangle$, we first need to introduce some new notations. Let 
\begin{equation}
A_i^{(j)}=\{ \g \in \Omega : (\s_i)_\g=j\}, \end{equation}
where $i \in N$ and $j \in F$. So that, any $\langle \s_i \rangle$ can be written as
\begin{equation*}
\langle \s_i \rangle=1 \cdot P(A_i^{(1)})+2 \cdot P(A_i^{(2)})+\cdots+q \cdot P(A_i^{(q)}).  \end{equation*}

\begin{dfnt} \label{trans}
For arbitrary permutation $\pi \in S_q$ let us define transformation $T_\pi:\Omega \to \Omega$ by the following way: for any $\g=(\s_1,\s_2,\cdots,\s_n)$ assume
\begin{equation*}
(T_\pi)_\g=\{\pi(\s_1),\pi(\s_2),\cdots,\pi(\s_n)\}. 
\end{equation*}
\end{dfnt}

\begin{rmk}\label{rmk1} For any transformation $T$ defined above, $P{((T_\pi)_\g)}=P(\g)$ for arbitrary $\g \in \Omega$, that is $P(A^{(\pi (j))}_i)=P(A^{(j)}_i)$ for any permutation $\pi \in S_q$ and any $j \in F$ (also for $j \in F^c$). 
\end{rmk}
The remark above follows from the fact that Kronecer's symbol only takes into account the similarity of spins. Since $T_\pi$ is a one-to-one transformation, it is also measure preserving. 

\begin{eg} If $n=4, q=3, \pi=(1\ 2\ 3)\to(2\ 1\ 3)$ and $\g=(1,1,3,2)$ then $(T_\pi)_\g=(2,2,3,1)$. Evidently $T_\pi$ is a one-to-one transformation $\Omega \to \Omega$. Notice that $P(1,1,3,2)=P(2,2,3,1)$ due to the usage of Kronecer's symbol in obtaining Gibbs probability. 
\end{eg}
It follows that for any $i \in N$
\begin{equation*}
P(A_i^{(1)}) = P(A_i^{(2)}) = \cdots = P(A_i^{(q)}) \end{equation*}
and since $P$ is a probabilistic measure, then for any $i \in N$ and $j \in F$,
\begin{equation*}
P(A_i^{(j)}) =  1/q. \end{equation*}
Therefore we have
\begin{equation}
\langle \s_i \rangle=(1 + 2 +\cdots+ q)/q = (q+1)/2 , \end{equation}
consequently enabling us to find $\s_\g^\prime$ for any $q$ values of spins by rewriting (\ref{cv}) as  
\begin{equation} 
\s_i^\prime=\s_i-(q+1)/2 , 
\end{equation}
which implies that $F^c=F-(q+1)/2$ hence the statements of Proposition \rf{prop1} follows. 
$\square$

Taking into account that changing the value of the spins from $F \to F^c$ does not affect the Hamiltonian as well as the Gibbs probability and also that the sum of spins in $F^c=0$, then for any $i \in N$,
\begin{equation}
\langle \s_i^\prime \rangle = 1/q\sum_{j \in F^c} j = 0. \end{equation} 

 
\section{Generalization of the First Griffiths' Inequality}

Now consider the following generalization of Potts model. Let $N$ denote the index set $\{1, 2, \cdots ,n \}$ and let $R$ be  a list of indices where any $i \in R$ would imply that $i \in N$. Then for any $R$, define
$$\s^R = \prod_{i \in R} \s_{i}^\prime \ \ \ \ (\s^\emptyset \equiv 1).$$
Let $R^{\prime} \subset N$ be the set of all elements in $R$. The difference between $R$ and $R^{\prime}$ is that $R$ may contain repeated indices while $R^{\prime}$ may not since it is a set.  When there is no repeated indices in $R$ we have $R^{\prime}=R$.

For each $A=\{i_1,\cdots,i_k\} \subset N$ where $k \ge 2$, let the extended real number $J_A \ge 0$ be given ($J_A = \infty$ is permitted), and define the Hamiltonian by
\begin{equation}\label{27}
H_\g=-\sum_{A \subset N} J_A\delta_{(\s^A)_\g}
\end{equation}
where the generalized Kronecer's symbol $\delta_{(\s^A)_\g}$ is
\begin{equation*}
\delta_{(\s^A)_\g}=\left\{
\begin{array}{ll}
1&  if \ (\s_{i_1})_\g=\cdots=(\s_{i_k})_\g \\
0&  otherwise.
\end{array}\right.
\end{equation*}
Let $x_A=\exp(J_A) \ge 1$. Define
\begin{equation}
Z_\g = exp (-H_\g) = \prod_{A \subset N}x_A \end{equation} 
which enables us to express the Gibbs probabilty on the space of configurations $\Omega$ as $P(\g)=Z_\g/Z$, where $Z=\sum_\g Z_\g$. Thus the expected value of any $(\s^R)_\g$ is given by
\begin{equation}\label{29}
\langle \s^{R} \rangle =\sum_{\g} (\s^R)_\g P(\g)= Z^{-1} \sum_\g (\s^R)_\g Z_\g. \end{equation} 

\begin{theorem} \label{thrm1} In probability space {\rm($\Omega,P$)} defined by {\rm(\rf{27})-(\rf{29})}, we have $\langle \s^{R} \rangle \ge 0$ for all $R$, a list where every $i \in R$ implies $i \in N$. \end{theorem}

Notice that since we allow repeated indices in $R$, then $R$ is comprised of odd and even groups of repeated indices. For example when $R=[1,2,3,3,4,4,4]$, it has three odd groups of indices which are [1], [2] and [4,4,4]. It also has one even group of indices [3,3]. Define odd groups in $R$ as $\theta_i$ when $i \in R$ is repeated an odd number of times. Similarly, define even groups in $R$ as 
$\epsilon_i$ when $i \in R$ is repeated an even number of times. So now, we can say that $R=[1,2,3,3,4,4,4]$ is comprised of $\theta_1$, $\theta_2$, $\epsilon_3$ and $\theta_4$ thus $R=[\theta_1,\theta_2,\epsilon_3,\theta_4]$.

Define $R^+ \subset \Omega$ to be a set of configurations where multiplication of all spins in $R^+$ gives a positive value and similarly $R^- \subset \Omega$ is a set of configurations where multiplication of all spins in $R^-$ gives a negative value. Also let $R^0 \subset \Omega$ be the set where the multiplication of spins are zero so that we have $\Omega=R^+ \cup R^-\cup R^0$. Note that we only have $R^0$ when $q$ is odd and it does not appear in $\langle \s^R \rangle$ (since $\s^R=0$), so we only need to consider $R^+$ and $R^-$. When $R$ is only comprised of even groups of repeated indices then $(\s^R)_\g=\prod_{i \in R}(\s^{\epsilon_i})_\g \ge 0$ thus $R^-=\emptyset$ and $\langle \s^{R} \rangle \ge 0$ since $P(\g) \ge 0$. 

Otherwise, consider cases where at least one odd group of repeated indices, $\theta_i$, exist in $R$ (which is true for all instances when $|R|$ is odd). In these cases, for each ${\g \in R^+}$ there exists a corresponding ${\g^\prime \in R^-}$ due to symmetrical properties of centered value variables in $F^c$ . Any element of either subsets can be transformed into a corresponding element of the other subset by choosing any $i \in R$ and multiplying $\s_i$ with $-1$ or simply multiplying any one of the $\theta_i$, $i \in R$ with $-1$.  Consequently, for these cases, if $q$ is even then $|R^+|=|R^-|=|\Omega|/2=q^{n}/2$ and if $q$ is odd, then $|R^+|=|R^-|=|\Omega-R^0|/2$, so that 
\begin{equation*}
\langle \s^R \rangle =\sum_{\g \in R^+} (\s^R)_\g P(\g) + \sum_{\g^\prime \in R^-} (\s^R)_{\g^\prime} P(\g^\prime).
\end{equation*}
Since $(\s^R)_{\g^\prime} = -(\s^R)_\g$, we can write
$$\langle \s^R \rangle =\sum_{\g \in R^+} (\s^R)_\g [ P(\g) - P(\g^\prime)]= Z^{-1} \sum_{\g \in R^+} (\s^R)_\g [Z_\g - Z_{\g^\prime}] .$$

When $|R|$ is odd, the one-to-one correspondence between $R^+$ and $R^-$ can also be obtained simply by multiplying ${(\s^R)}_\g$ with $-1$. By this way the difference of spins are perserved hence the Gibbs measure is preserved, consequently $Z_\g=Z_{\g^\prime}$ for any $\g$. In other words for $|R|$ odd, $T_\pi: R^+ \to R^-$, where $T_\pi$ is stated in Definition \rf{trans}. Hence, since $(\s^R)_\g=-(\s^R)_{\g^\prime}$ and $Z_\g=Z_{\g^\prime}$ for all odd $|R|$, $\langle \s^R \rangle = Z^{-1} \sum_{\g \in R^+} (\s^R)_\g [Z_\g - Z_{\g^\prime}] = 0$ and Theorem \rf{thrm1} stands.


Let $A \subset \Omega$ be a set of configurations, and then define
$$\zeta(R,A)= \sum_{\g \in A} (\s^R)_\g Z_\g.$$
For brevity if $A=\Omega$ let $\zeta(R)=\zeta(R, \Omega)$. Thus we can write
$$\zeta(R)= \sum_{\g \in \Omega} (\s^R)_\g Z_\g = Z \cdot \langle \s^R \rangle ,$$
and when $R^+=\{\g \in \Omega:\s^R>0\}$ and $R^-=\{\g \in \Omega:\s^R<0\}$, we have
$$\zeta(R)=\zeta(R,R^+)+\zeta(R,R^-)= \sum_{\g \in R^+} (\s^R)_\g Z_\g + \sum_{\g^\prime \in R^-} (\s^R)_{\g^\prime} Z_{\g^\prime}=\sum_{\g \in R^+} (\s^R)_\g [Z_\g - Z_{\g^\prime}].$$ 

Let $B \subset N$, where $B^{(1)}=\{\g \in \Omega:\delta_{{(\s^B)}_\g} = 1\}$ and $B^{(0)}=\{\g \in \Omega:\delta_{{(\s^B)}_\g} = 0\}$, then since $\Omega=B^{(1)} \cup B^{(0)}$, 
$$\zeta(R)= \sum_{\g \in B^{(1)}} (\s^R)_\g Z_\g + \sum_{\g \in B^{(0)}} (\s^R)_{\g} Z_{\g}=\zeta(R,B^{(1)})+\zeta(R,B^{(0)}).$$
Similarly, let $R^+B^{(1)}=\{\g \in \Omega:\s^R>0 \ and \ \delta_{{(\s^B)}_\g} = 1\}$, $R^-B^{(1)}=\{\g \in \Omega:\s^R<0 \ and \ \delta_{{(\s^B)}_\g} = 1\}$, $R^+B^{(0)}=\{\g \in \Omega:\s^R>0 \ and \ \delta_{{(\s^B)}_\g} = 0\}$ and $R^-B^{(0)}=\{\g \in \Omega:\s^R<0 \ and \ \delta_{{(\s^B)}_\g} = 0\}$ for any $B \subset N$, we can write
$$\zeta(R,\Omega)=\sum_{\g \in R^+B^{(1)}}(\s^R)_\g Z_\g +  \sum_{\g^\prime \in R^-B^{(1)}}(\s^R)_{\g^\prime}  Z_{\g^\prime} + \sum_{\g \in R^+B^{(0)}}(\s^R)_\g Z_\g +  \sum_{\g^\prime \in R^-B^{(0)}}(\s^R)_{\g^\prime}  Z_{\g^\prime}$$
$$=\zeta(R,R^+B^{(1)})+\zeta(R,R^-B^{(1)})+\zeta(R,R^+B^{(0)})+\zeta(R,R^-B^{(0)}).$$


For cases where $|R|$ is even and $(\s^R)_\g \not= \prod_{i \in R}(\s^{\epsilon_i})_\g$ (there exists $\theta_i$, $i \in R$), we seek to prove by induction on $s$, the number of $J_A > 0$. To prove $\langle \s^R \rangle \ge 0$ we only need to prove that $\zeta(R) \ge 0$ since $\zeta(R)=Z \cdot \langle \s^R \rangle$ and $Z>0$. For $s=0$ we have $Z_\g=Z_{\g^\prime}=1$, thus 
$$\zeta(R)=\sum_{\g \in R^+} (\s^R)_\g [Z_\g - Z_{\g^\prime}]=\sum_{\g \in R^+} (\s^R)_\g [1 - 1]=0$$
and Theorem \rf{thrm1} is satisfied. Note that $P(\g)=1/Z$ for all $\g$ hence we have uniform measure which renders $\langle \s^R \rangle = 0$ since $\sum_{\g \in R}(\s^R)_\g=0$ due to centered value properties. 

Let $\zeta_s(R)$ be $\zeta(R)$ for any $s$ number of nonzero existing interactions. Assume $\zeta_s(R) \ge 0$ for all $s \leq k$ such that for any $s$ we add $J_{B_s} >0$. Then for $s=k+1$ let $J_{B_{k+1}} > 0$ be the additional interaction. Since we know that $x_{B_{k+1}}$ will only multiply all the terms in $B_{k+1}^{(1)}$ where $B_{k+1}^{(1)}=\{\g \in \Omega:\delta_{{(\s^{B_{k+1}})}_\g} = 1\}$, the terms in $B_{k+1}^{(0)}$ ($B_{k+1}^{(0)}=\{\g \in \Omega:\delta_{{(\s^{B_{k+1}})}_\g} = 0\}$) remains the same as it was in $s=k$. Thus we have
$$\zeta_{k+1}(R)=\zeta_{k+1}(R,B_{k+1}^{(1)})+\zeta_{k+1}(R,B_{k+1}^{(0)})=x_{B_{k+1}} \cdot \zeta_k (R,B_{k+1}^{(1)})+\zeta_k(R,B_{k+1}^{(0)}).$$ 

By induction hypothesis we have $\zeta_k(R) \ge 0$, and if we have $\zeta_k(R,B_{k+1}^{(1)}) \ge 0$ we shall be able to write
$$\zeta_{k+1}(R)=x_{B_{k+1}} \cdot \zeta_k (R,B_{k+1}^{(1)})+\zeta_k(R,B_{k+1}^{(0)}) > \zeta_k (R,B_{k+1}^{(1)})+\zeta_k(R,B_{k+1}^{(0)})= \zeta_k (R) \ge 0.$$
since $x_{B_{k+1}} > 1$ and it is multiplied with a positive sum. Thus given $\zeta_k(R,B_{k+1}^{(1)}) \ge 0$, we have $\langle \s^R \rangle \ge 0$ for any $n$ number of vertices, $q$ number of spins and $R$ in which any $i \in R$ is also in $N$.

\begin{lemma}\label{lemma2}

Let $\zeta^N_s(R)$ be $\zeta(R)$ where $s$ is the number of nonzero $J_A$, $N=\{1,\cdots,n\}$ is the set of $n$ vertices and $R$ is a list where $i \in R$ implies $i \in N$. Given that $\zeta^N_s(R) \ge 0$ for any $n$ vertices and $q$ number of spins then we have $\zeta^N_s(R,B^{(1)}) \ge 0$ where $B \subset N$ and $B^{(1)}=\{\g \in \Omega:\delta_{{(\s^B)}_\g} = 1\}$.
 
\end{lemma}

\noindent {\it Proof.} Let $B=\{b_1,b_2 \cdots b_m\} \subset N$. For $\g \in B^{(1)}$, the spins are always similar for all its vertices, such that $\s_{b_1} = \s_{b_2}= \cdots =\s_{b_m}$, thus we seek to treat $B$ as a single vertex, say $b_1$. Firstly get $B \cap A$ for all existing $J_A$'s, if $B \cap A = \emptyset$, then $x_A$ is left as it is, but if $B \cap A \not= \emptyset$ then we shall do some alterations. 

For $A$'s where $B \cap A \not= \emptyset$ let $C_A = A - (B \cap A)$. If there exist $A$'s with similar $C_A$, then we seek to group it together. Let $C =C_A \cup b_1$ where $b_1$ is the first element of $B$ then we set $x_C^*=\prod x_A$ for all $A$'s with similar $C_A$, to represent them in a group as a single interaction. We can do this because they will always appear together in $\zeta_s (R,B^{(1)})$. Thus if there exist similar $C_A$'s for different $A$'s, the number of existing interaction is reduced but the remaining interaction has a larger size which does not matter since $J_A$ can even be $\infty$. If $C_A = \emptyset$ then the $x_{b_1}^*$ group if comprised of all of $A \subset B$. This group will appear in every in term in $\zeta (R,B^{(1)})$ due to the fact that all spins in $B$ are similar for $\g \in B^{(1)}$ . 

After replacing all the terms where $B \cap A \not= \emptyset$ with its corresponding $x_C^*$, then we will see that we have
\begin{equation}\label{31}
\zeta^N_k(R,B^{(1)}) = x_{b_1}^* \cdot \zeta^{N^*}_s(R^*)
\end{equation}
where $s \le k$, $N^*=(N-B)\cup b_1$ and $R^*$ is obtain by simply replacing any $i \in R$ which is also in $B$ with $b_1$ (if none of $i \in R$ is in $B$ then $R^*=R$). $(\s^R)_\g=(\s^{R^*})_\g$ since the spins in $B$ are all similar, we are simply renaming the vertices. A simple example is that initially we have $B=\{2,3\}$ thus $b_1=2$ and $R=[1,2,3,4]$ in $N=\{1,2,3,4\}$ then we can see that the $\g \in B^{(1)}$ is exactly similar to $\g \in \Omega$ for cases where $R^*=[1,2,2,4]$ in $N^*=\{1,2,4\}$ which can be modified by renaming index 4 as index 3 and then it can be obtained just like in the case where $R=[1,2,2,3]$ in $N=\{1,2,3\}$.

We can find $\zeta^{N^*}_s(R^*)$ exactly the same way we obtain $\zeta^{N}_s(R)$ where $N=\{1,\cdots,n^*\}$, $n^*=n-|B|+1$ and $R^*$ is transformed accordingly to a new $R$ where $|R^*|=|R|$, the difference is only that the vertices have different position, but because any interactions are accounted for, this does not really matter since the existence and size of interactions does not depends on the vertices being neighbours or not. Now that we know  $x_{b_1}^* \ge 1$, $\zeta^{N^*}_s(R^*) \ge 0$ (since $\zeta^{N}_s(R) \ge 0$ for any  $n$ including $n^*$ and any $R$ where $i \in R$ implies $i \in N$), then we have $\zeta^N_k(R,B^{(1)}) \ge 0$. $\square$

\begin{eg}

In this example we seek to illustrate Lemma 2. Let $N=\{1,2,3\}$, $q=3$, $R= [1,3]$ and $B= \{1,2\}$. The only possible interactions are $J_{12}$, $J_{13}$, $J_{23}$, $J_{123}$. Assume only $x_{12}=1$ hence $s=3$. We also have
$$B^{(1)}= \{(-1,-1,-1),(-1,-1,1),(1,1,-1),(1,1,1)\} \  and \ \zeta(R,B^{(1)})= 2(x_{13}x_{23}x_{123}-1).$$
Since $C_{13}=\{3\}$, $C_{23}=\{3\}$ and $C_{123}=\{3\}$, assign $x_{13}^*=x_{13}x_{23}x_{123}$. Replace it in $\zeta(R,B^{(1)})$, we have 
$$\zeta(R,B^{(1)})= 2(x_{13}^*-1).$$ 
Consequently we have only one existing interaction, $x_{13}^*$. Note that if $x_{12}\ge 1$ than it will be $\zeta(R,B^{(1)})= 2x_1^*(x_{13}^*-1)$ where $x_1^*=x_{12}$.
Now let $N^*=(N-B)\cup b_1= \{1,3\}$ thus the only possible interaction here is $x_{13}^*$. $R^*=R$ since the only similar elements of $R$ and $B$ is $b_1$. The set of possible configuration are 
$$\{(-1,-1),(-1,0),(-1,1),(1,-1),(1,0),(1,1)\}$$ thus we have
$$R^{*+}= \{(-1,-1),(1,1)\} \ and \ R^{*-}=\{(-1,1),(-1,1)\},$$
$$\zeta(R^*)=2(x_{13}^*-1)$$ 
thus we see that equation {\rm(\rf{31})} is verified. Note that $N^*=\{1,3\}$ can be treated like $N=\{1,2\}$ with $n=2$ if we change vertex 3 to vertex 2 {\rm(}thus $R^*= [1,3]$ becomes $R= [1,2]${\rm )}. For $N=\{1,2\}$ we will get $\zeta(R)=2(x_{12}-1)$ where $x_{12}=x_{13}^*$.

\end{eg}

Since we can just replace $B$ in Lemma 2 by $B_{k+1}$ due to the fact that both are subsets of $N$, thus by Lemma 2 we have $\zeta_k(R,B_{k+1}^{(1)}) \ge 0$ when $\zeta_k(R) \ge 0$, hence we have proven $\langle \s^R \rangle \ge 0$ for any $n$ number of vertices, $q$ number of spins and $R$ where $i \in R$ implies $i \in N$.


\section{Generalization of the Second Griffiths' Inequality}

Similar to the definition of $R$, let $S$ be  a list of indices where any $i \in S$ would imply that $i \in N$. Then for any $S$, define
$$\s^S = \prod_{i \in S} \s_{i}^\prime \ \ \ \ (\s^\emptyset \equiv 1).$$
We also let $S^{\prime} \subset N$ be the set of all elements in $S$. Subsequently we have $RS=[R,S]$ thus $RS^\prime=R^{\prime} \cup S^{\prime}$. $RS$ is a list comprised of $R$ and $S$, so for any $RS$, 
$$\s^{RS} = \s^R\s^S= \prod_{i \in R} \s_{i}^\prime \prod_{i \in S} \s_{i}^\prime.$$

\begin{theorem} \label{thrm2} In probability space \rm{($\Omega,P$)} defined by \rm{(\rf{27})-(\rf{29})}, we have $\langle \s^{R}\s^{S} \rangle-\langle \s^{R} \rangle\langle \s^{S} \rangle \ge 0$ for all $R^\prime,S^\prime \subset N$. \end{theorem}
 
\noindent Note that, when either $|R|$ or $|S|$ is odd then $|RS|$ is also odd thus we have 
$$\langle \s^{RS} \rangle-\langle \s^{R} \rangle\langle \s^{S}\rangle= 0-0 =0 $$
which fulfills Theorem 2. If both $|R|$ and $|S|$ are odd then $|RS|$ is even and we have
$$\langle \s^{RS} \rangle-\langle \s^{R} \rangle\langle \s^{S}\rangle= \langle \s^{RS} \rangle-0 =\langle \s^{RS} \rangle \ge 0$$
which also fulfills Theorem 2.

To complete prove for Theorem 2 we only need to prove for cases where $|R|$ and $|S|$ is even thus $|RS|$ is also even. We seek to prove by induction on $s$, the number of $J_A > 0$. To prove $\langle \s^{RS} \rangle-\langle \s^{R} \rangle\langle \s^{S}\rangle \ge 0$ we only need to prove that $Z \cdot \zeta(RS)-\zeta(S) \cdot \zeta(S) \ge 0$ since $Z>0$ and
$$\langle \s^{R}\s^{S} \rangle-\langle \s^{R} \rangle\langle \s^{S}\rangle =
\langle \s^{RS} \rangle-\langle \s^{R} \rangle\langle \s^{S}\rangle=\frac{\zeta(RS)}{Z}-\frac{\zeta(R)}{Z} \cdot \frac{\zeta(S)}{Z}=\frac{Z \cdot \zeta(RS)-\zeta(S) \cdot \zeta(S)}{Z^2}$$
due to the fact that $\zeta(R)=Z \cdot \langle \s^R \rangle$ . 

For $s=0$ we have $Z_\g=Z_{\g^\prime}=1$ for any $\g$, thus as long as there is correspondence between $R^+$ and $R^-$ we have $\langle \s^{R} \rangle=0$. But we only have the correspondence when there is at least one odd group of repeated indices, $\theta_i$ in $R$. This is also true for $S$ and $RS$. If there exist an odd group of indices in $R$, $S$ or both of them ,then we have 
$$\langle \s^{RS} \rangle-\langle \s^{R} \rangle\langle \s^{S}\rangle= \langle \s^{RS} \rangle-0 =\langle \s^{RS} \rangle \ge 0.$$

When $R$ and $S$ are comprised of only even group of indices we do not have the correspondence. Let $\epsilon_i^R$ be $\epsilon_i$ given $R$ and $\theta_i^R$ be $\theta_i$ given $R$, for any $R$ where $i \in R$ implies that $i \in N$. In such cases $(\s^R)_\g=\prod_{i \in R}(\s^{\epsilon_i^R})_\g \ge 0$, $(\s^S)_\g=\prod_{i \in S}(\s^{\epsilon_i^S})_\g \ge 0$ and $(\s^{RS})_\g=\prod_{i \in RS}(\s^{\epsilon_i^{RS}})_\g \ge 0$. Thus $\langle \s^{R} \rangle \ge 0$, $\langle \s^{S} \rangle \ge 0$ and $\langle \s^{RS} \rangle \ge 0$. Examples are $R=[1,1,2,2]$ and $S=[3,3]$ thus $RS=[1,1,2,2,3,3]$. Since $Z_\g=1$ we have
$$\zeta(R)=\sum_{\g}(\s^R)_\g=\sum_{\g}\prod_{i \in R}(\s^{\epsilon_i^R})_\g=q^{|N-R^\prime|} \cdot\prod_{i \in R}\sum_{j \in F^c}j^{|\epsilon_i^R|}$$
and similarly for $S$
$$\zeta(S)=\sum_{\g}(\s^S)_\g=\sum_{\g}\prod_{i \in S}(\s^{\epsilon_i^S})_\g=q^{|N-S^\prime|} \cdot\prod_{i \in S}\sum_{j \in F^c}j^{|\epsilon_i^S|}.$$
Also due to the fact that $Z_\g=1$ we have $Z=|\Omega|=q^{|N|}=q^n$ and 
$$\zeta(RS)=\sum_{\g}(\s^{RS})_\g=\sum_{\g}\prod_{i \in RS}(\s^{\epsilon_i^{RS}})_\g=q^{|N-RS^\prime|} \cdot\prod_{i \in RS}\sum_{j \in F^c}j^{|\epsilon_i^{RS}|}.$$
Consequently we have 
$$Z \cdot \zeta(RS)-\zeta(S) \cdot \zeta(S)=q^nq^{|N-RS^\prime|} \cdot\prod_{i \in RS}\sum_{j \in F^c}j^{|\epsilon_i^{RS}|}-[q^{|N-R^\prime|} \cdot\prod_{i \in R}\sum_{j \in F^c}j^{|\epsilon_i^R|}][q^{|N-S^\prime|} \cdot\prod_{i \in S}\sum_{j \in F^c}j^{|\epsilon_i^S|}].$$

If $R^\prime \cap S^\prime = \emptyset$ then we have
$$q^nq^{|N-RS^\prime|} =q^{|N-R^\prime|}q^{|N-S^\prime|} \ \ \ \text{and} \ \ \ 
\prod_{i \in RS}\sum_{j \in F^c}j^{|\epsilon_i^{RS}|}=\prod_{i \in R}\sum_{j \in F^c}j^{|\epsilon_i^R|} \cdot \prod_{i \in S}\sum_{j \in F^c}j^{|\epsilon_i^S|}$$ 
thus we can write
$$Z \cdot \zeta(RS)=q^n q^{|N-RS^\prime|} \cdot\prod_{i \in RS}\sum_{j \in F^c}j^{|\epsilon_i^{RS}|}=q^{|N-R^\prime|}q^{|N-S^\prime|} \cdot \prod_{i \in R}\sum_{j \in F^c}j^{|\epsilon_i^R|} \cdot \prod_{i \in S}\sum_{j \in F^c}j^{|\epsilon_i^S|}$$ 
$$=q^{|N-R^\prime|} \cdot\prod_{i \in R} \sum_{j \in F^c}j^{|\epsilon_i^R|} \cdot q^{|N-S^\prime|} \cdot\prod_{i \in S} \sum_{j \in F^c}j^{|\epsilon_i^S|}= \zeta(R) \cdot \zeta(S)$$
so that $Z \cdot \zeta(RS)-\zeta(R) \cdot \zeta(S)=0$ thus $\langle \s^{RS} \rangle-\langle \s^{R} \rangle\langle \s^{S}\rangle=0$.

But if $R^\prime \cap S^\prime \not= \emptyset$ then $q^nq^{|N-RS^\prime|} =q^{|N-R^\prime|}q^{|N-S^\prime|}q^{|R^\prime \cap S^\prime|}$ and only for $i \not\in (R^\prime \cap S^\prime )$ do we have
$$\prod_{i \in RS^\prime-(R^\prime \cap S^\prime )} \sum_{j \in F^c}j^{|\epsilon_i^{RS}|}=[\prod_{i \in R^\prime-(R^\prime \cap S^\prime )} \sum_{j \in F^c}j^{|\epsilon_i^R|}] \cdot [\prod_{i \in S^\prime-(R^\prime \cap S^\prime )} \sum_{j \in F^c}j^{|\epsilon_i^S|}] $$
and we can write

$$Z \cdot \zeta(RS)-\zeta(R) \cdot \zeta(S)=q^nq^{|N-RS^\prime|} \cdot\prod_{i \in RS}\sum_{j \in F^c}j^{|\epsilon_i^{RS}|}-[q^{|N-R^\prime|} \cdot\prod_{i \in R}\sum_{j \in F^c}j^{|\epsilon_i^R|}][q^{|N-S^\prime|} \cdot\prod_{i \in S}\sum_{j \in F^c}j^{|\epsilon_i^S|}]$$

$$=q^{|N-R^\prime|}q^{|N-S^\prime|}q^{|R^\prime \cap S^\prime|}\cdot\prod_{i \in RS}\sum_{j \in F^c}j^{|\epsilon_i^{RS}|}-q^{|N-R^\prime|}q^{|N-S^\prime|} \cdot \prod_{i \in R}  \sum_{j \in F^c}j^{|\epsilon_i^R|} \cdot \prod_{i \in S} \sum_{j \in F^c}j^{|\epsilon_i^S|}$$

$$=q^{|N-R^\prime|+|N-S^\prime|}\prod_{\stackrel{i \in R^\prime \ and \ }{i \not\in (R^\prime \cap S^\prime )}} \sum j^{|\epsilon_i^{RS}|}[q^{|R^\prime \cap S^\prime|}  \prod_{i \in (R^\prime \cap S^\prime )} \sum j^{|\epsilon_i^{RS}|} - \prod_{i \in (R^\prime \cap S^\prime )} \sum j^{|\epsilon_i^R|}  \prod_{i \in (R^\prime \cap S^\prime )} \sum j^{|\epsilon_i^S|}]
$$

So now we only need to prove that
$$q^{|R^\prime \cap S^\prime|} \cdot \prod_{i \in (R^\prime \cap S^\prime )} \sum j^{|\epsilon_i^{RS}|} - \prod_{i \in (R^\prime \cap S^\prime )} \sum j^{|\epsilon_i^R|} \cdot \prod_{i \in (R^\prime \cap S^\prime )} \sum j^{|\epsilon_i^S|} \ge 0.$$
We can do that by proving that for each $i \in (R^\prime \cap S^\prime)$ we have 
$$q\sum_{j \in F^c}j^{|\epsilon_i^{RS}|}-\sum_{j \in F^c}j^{|\epsilon_i^{R}|}\sum_{j \in F^c}j^{|\epsilon_i^{S}|}$$ 
since
$$q^{|R^\prime \cap S^\prime|}  \cdot  \prod_{i \in (R^\prime \cap S^\prime )} \sum j^{|\epsilon_i^{RS}|} - \prod_{i \in (R^\prime \cap S^\prime )} \sum j^{|\epsilon_i^R|}  \cdot  \prod_{i \in (R^\prime \cap S^\prime )} \sum j^{|\epsilon_i^S|}$$
$$ = \prod_{i \in (R^\prime \cap S^\prime)}[q\sum_{j \in F^c}j^{|\epsilon_i^{RS}|}]-\prod_{i \in (R^\prime \cap S^\prime)}[\sum_{j \in F^c}j^{|\epsilon_i^{R}|}\sum_{j \in F^c}j^{|\epsilon_i^{S}|}].$$

Let $\xi_q=q\sum_{j \in F^c}j^{|\epsilon_i^{RS}|}-\sum_{j \in F^c}j^{|\epsilon_i^{R}|}\sum_{j \in F^c}j^{|\epsilon_i^{S}|}$ for any $i \in (R^\prime \cap S^\prime)$ and $q$. Note that we have $|\epsilon_i^{RS}|=|\epsilon_i^{R}+\epsilon_i^{S}|=|\epsilon_i^{R}|+|\epsilon_i^{S}|$ for any $R$, $S$ and $RS$. When $q=2$, $F^c=\{-1/2,1/2\}$ and 
$$\xi_2=2 \cdot [(-1/2)^{|\epsilon_i^{RS}|}+(1/2)^{|\epsilon_i^{RS}|}] -[(-1/2)^{|\epsilon_i^{R}|}+(1/2)^{|\epsilon_i^{R}|}]\cdot[(-1/2)^{|\epsilon_i^{S}|}+(1/2)^{|\epsilon_i^{S}|}]$$
$$=2 \cdot 2(1/2)^{|\epsilon_i^{RS}|}-2(1/2)^{|\epsilon_i^{R}|}\cdot2(1/2)^{|\epsilon_i^{S}|}
=2 \cdot 2(1/2)^{|\epsilon_i^{R}+\epsilon_i^{S}|}-2(1/2)^{|\epsilon_i^{R}|}\cdot2(1/2)^{|\epsilon_i^{S}|}$$
$$=2 \cdot 2(1/2)^{|\epsilon_i^{R}|}(1/2)^{|\epsilon_i^{S}|}-2(1/2)^{|\epsilon_i^{R}|}\cdot2(1/2)^{|\epsilon_i^{S}|}=0.$$
If $q=3$ then $F^c=\{-1,0,1\}$ and 
$$\xi_3=3 \cdot [(-1)^{|\epsilon_i^{RS}|}+0^{|\epsilon_i^{RS}|}+1^{|\epsilon_i^{RS}|}] -[(-1)^{|\epsilon_i^{R}|}+0^{|\epsilon_i^{R}|}+1^{|\epsilon_i^{R}|}]\cdot[(-1)^{|\epsilon_i^{S}|}+0^{|\epsilon_i^{S}|}+1^{|\epsilon_i^{S}|}]$$
$$=3 \cdot 2[1^{|\epsilon_i^{RS}|}] -2[1^{|\epsilon_i^{R}|}]\cdot2[1^{|\epsilon_i^{S}|}]=3 \cdot 2-2 \cdot 2= 2 >0.$$
Now to find a general formula for any $\xi_q$, by obtaining $\xi_{q+2}-\xi_q$, using the fact that
$$\xi_{q+2}=(q+2)(\sum_{j \in F^c}j^{|\epsilon_i^{RS}|}+2(\frac{q+1}{2})^{|\epsilon_i^{RS}|})-(\sum_{j \in F^c}j^{|\epsilon_i^{R}|}+2(\frac{q+1}{2})^{|\epsilon_i^{R}|})(\sum_{j \in F^c}j^{|\epsilon_i^{S}|}+2(\frac{q+1}{2})^{|\epsilon_i^{S}|})$$ 
$$\ \ \ \ \ \text{and} \ \ \ \ \ \xi_q=q\sum_{j \in F^c}j^{|\epsilon_i^{RS}|}-\sum_{j \in F^c}j^{|\epsilon_i^{R}|}\sum_{j \in F^c}j^{|\epsilon_i^{S}|}.$$
As a result we have
\begin{eqnarray*}
\xi_{q+2}-\xi_q&=&2\sum_{j \in F^c}j^{|\epsilon_i^{RS}|}+2(q+2)(\frac{q+1}{2})^{|\epsilon_i^{RS}|}-2(\frac{q+1}{2})^{|\epsilon_i^{R}|}\sum_{j \in F^c}j^{|\epsilon_i^{S}|} \\
& &-2(\frac{q+1}{2})^{|\epsilon_i^{S}|}\sum_{j \in F^c}j^{|\epsilon_i^{R}|}-4(\frac{q+1}{2})^{|\epsilon_i^{R}|}(\frac{q+1}{2})^{|\epsilon_i^{S}|}.
\end{eqnarray*}
We know that $|\epsilon_i^{RS}|=|\epsilon_i^{R}+\epsilon_i^{S}|$, then
$$\xi_{q+2}-\xi_q=2\sum_{j \in F^c}j^{|\epsilon_i^{RS}|}+2(q+2-2)(\frac{q+1}{2})^{|\epsilon_i^{RS}|}-2(\frac{q+1}{2})^{|\epsilon_i^{R}|}\sum_{j \in F^c}j^{|\epsilon_i^{S}|}-2(\frac{q+1}{2})^{|\epsilon_i^{S}|}\sum_{j \in F^c}j^{|\epsilon_i^{R}|}$$
$$=2[q(\frac{q+1}{2})^{|\epsilon_i^{RS}|}-(\frac{q+1}{2})^{|\epsilon_i^{R}|}\sum_{j \in F^c}j^{|\epsilon_i^{S}|}-(\frac{q+1}{2})^{|\epsilon_i^{S}|}\sum_{j \in F^c}j^{|\epsilon_i^{R}|}+\sum_{j \in F^c}j^{|\epsilon_i^{RS}|}]$$
$$=2\sum_{j \in F^c}[(\frac{q+1}{2})^{|\epsilon_i^{R}|}-j^{|\epsilon_i^{R}|}][(\frac{q+1}{2})^{|\epsilon_i^{S}|}-j^{|\epsilon_i^{S}|}]\ge 0$$
since the largest $j$ is $\frac{q-1}{2}$ thus $[(\frac{q+1}{2})^{|\epsilon_i^{R}|}-j^{|\epsilon_i^{R}|}] > 0$ and $[(\frac{q+1}{2})^{|\epsilon_i^{S}|}-j^{|\epsilon_i^{S}|}] > 0$.

We have established that $\xi_2=0$, $\xi_3=2$ and
$$\xi_{q+2}=\xi_{q}+2\sum_{j \in F^c}[(\frac{q+1}{2})^{|\epsilon_i^{R}|}-j^{|\epsilon_i^{R}|}][(\frac{q+1}{2})^{|\epsilon_i^{S}|}-j^{|\epsilon_i^{S}|}]$$
so that $\xi_q \ge 0$ for any $q \ge 2$. Therefore when $s=0$ we have $Z \cdot \zeta(RS)-\zeta(S) \cdot \zeta(S) \ge 0$ thus $\langle \s^{RS} \rangle-\langle \s^{R} \rangle\langle \s^{S}\rangle \ge 0$ for any $R$ and $S$.

Let $Z \cdot \zeta(RS)-\zeta(S) \cdot \zeta(S) \ge 0$ for any $s \le k$, number of nonzero $J_A$'s. Consider the case when $s=k+1$, and the added interaction is $J_{B_{k+1}}$. Let $x=x_{B_{k+1}}$ for brevity. Then define $Z^{(1)}=\sum_{\g \in B^{(1)}_{k+1}}Z_\g$ and $Z^{(0)}=\sum_{\g \in B^{(0)}_{k+1}}Z_\g$ so that $$Z=Z^{(1)} \cdot x+Z^{(0)}.$$ 
We also have these equations:
$$\zeta(RS)=\zeta(RS,B^{(1)}_{k+1}) \cdot x + \zeta(RS,B^{(0)}_{k+1}).$$
$$\zeta(R)=\zeta(R,B^{(1)}_{k+1}) \cdot x + \zeta(R,B^{(0)}_{k+1}).$$ 
$$\zeta(S)=\zeta(S,B^{(1)}_{k+1}) \cdot x +\zeta(S,B^{(0)}_{k+1}).$$ 


Due to the fact that
\begin{eqnarray*}
Z \cdot \zeta(RS)- \zeta(R) \cdot \zeta(S)&=&[Z^{(1)} \cdot x+Z^{(0)}][\zeta(RS,B^{(1)}_{k+1}) \cdot x + \zeta(RS,B^{(0)}_{k+1})] \\
&&-[\zeta(R,B^{(1)}_{k+1}) \cdot x + \zeta(R,B^{(0)}_{k+1})][\zeta(S,B^{(1)}_{k+1}) \cdot x + \zeta(S,B^{(0)}_{k+1})],
\end{eqnarray*}
we can write $Z \cdot \zeta(RS)- \zeta(R) \cdot \zeta(S)= Ux^2+Vx+W$ where 
$$U=Z^{(1)} \cdot \zeta(RS,B^{(1)}_{k+1})- \zeta(R,B^{(1)}_{k+1}) \cdot \zeta(S,B^{(1)}_{k+1}),$$
$$V=Z^{(1)} \cdot \zeta(RS,B^{(0)}_{k+1})+Z^{(0)} \cdot \zeta(RS,B^{(1)}_{k+1})
- \zeta(R,B^{(0)}_{k+1}) \cdot \zeta(S,B^{(1)}_{k+1})- \zeta(R,B^{(1)}_{k+1}) \cdot \zeta(S,B^{(0)}_{k+1})$$
$$\ \ \text{and} \ \ W=Z^{(0)} \cdot \zeta(RS,B^{(0)}_{k+1})- \zeta(R,B^{(0)}_{k+1}) \cdot \zeta(S,B^{(0)}_{k+1}).$$

Since $Z^{(1)} \cdot \zeta(RS,B^{(1)}_{k+1})- \zeta(R,B^{(1)}_{k+1}) \cdot \zeta(S,B^{(1)}_{k+1})=x^*_{b_1}[Z^* \cdot \zeta(RS^*)- \zeta(R^*) \cdot \zeta(S^*)]$ by Lemma 1 and $Z^* \cdot \zeta(RS^*)- \zeta(R^*) \cdot \zeta(S^*)\ge 0$ by induction hypothesis, we have $U \ge 0$ so that $Ux^2+Vx+W$ is a quadratic function with a minimum value. By definition we know that 
\begin{eqnarray*}
2U+V&=&2[Z^{(1)} \cdot \zeta(RS,B^{(1)}_{k+1})- \zeta(R,B^{(1)}_{k+1}) \cdot \zeta(S,B^{(1)}_{k+1})] + Z^{(1)} \cdot \zeta(RS,B^{(0)}_{k+1})\\
&&+Z^{(0)} \cdot \zeta(RS,B^{(1)}_{k+1})
- \zeta(R,B^{(0)}_{k+1}) \cdot \zeta(S,B^{(1)}_{k+1})- \zeta(R,B^{(1)}_{k+1}) \cdot \zeta(S,B^{(0)}_{k+1})\\
&=& Z^{(1)} \cdot \zeta(RS)+Z \cdot \zeta(RS,B^{(1)}_{k+1})
- \zeta(R) \cdot \zeta(S,B^{(1)}_{k+1})- \zeta(R,B^{(1)}_{k+1}) \cdot \zeta(S).
\end{eqnarray*}
It is easy to verify that since we have $U \ge 0$ and $U+V+W \ge 0$ (by induction hypothesis) we also have $2U+V \ge 0$. 

By differentiating $Ux^2+Vx+W$ where $x>1$ , we obtain 
$$\frac{d(Ux^2+Vx+W)}{dx}=2Ux+V$$
Given that we have $2U+V \ge 0$ , we can conclude that $Ux^2+Vx+W$ is an increasing function for any $x \ge 1$.  When $x =1$, we have $Ux^2+Vx+W=U+V+W \ge 0$ , so now we know that $Ux^2+Vx+W$ is always positive. Hence we have $Z \cdot \zeta(RS)- \zeta(R) \cdot \zeta(S) \ge 0$ for any $x \ge 1$. Consequently $\langle \s^{R}\s^{S} \rangle-\langle \s^{R} \rangle\langle \s^{S}\rangle \ge 0$ for any $R$ and $S$ hence Theorem 2 is proven.

{\bf Acknowledgement.} The work is partly supported by the International Islamic University Malaysia short term research grant.

\end{document}